\newif\ifcomment
\newif\ifdraft
\newif\iflatexdiff
\def\dvers{v1.1}
\def\dtitle{Jet-associated deuteron production\\ in pp collisions at $\mathbf{\sqrt{s}=13}$ TeV}
\def\stitle{Jet-associated deuteron production in pp at 13 TeV}
\documentclass[ALICE,manyauthors]{cernphprep}
\usepackage[comma,square,numbers,sort&compress]{natbib}
\usepackage{hyperref}
\usepackage{graphicx}  
\usepackage{dcolumn}   
\usepackage{bm}        
\usepackage{amssymb}   
\usepackage{amsfonts}
\usepackage{graphics}
\usepackage{grffile}   
\usepackage{epsfig}
\usepackage[loose]{units}
\usepackage[usenames,dvipsnames]{xcolor}
\usepackage[normalem]{ulem} 
\usepackage{color}
\usepackage[utf8]{inputenc}
\usepackage[T1]{fontenc}
\usepackage{subcaption}
\DeclareSymbolFont{newfont}{OML}{cmm}{m}{it}
\DeclareMathSymbol{\Epsilon}{3}{newfont}{15}
\usepackage{verbatim}   
\usepackage{booktabs}
\usepackage{multicol}
\iflatexdiff
\RequirePackage{color}\definecolor{RED}{rgb}{1,0,0}\definecolor{BLUE}{rgb}{0,0,1}

\fi

\newcommand{\Yd}           {\ensuremath{Y_{\rm deuteron}}}
\newcommand{\dphi}         {\ensuremath{\Delta \varphi}}

\newcommand{\pp}           {pp}

\newcommand{\PbPb}         {\mbox{Pb--Pb}}

\newcommand{\pt}           {\ensuremath{p_{\mathrm{T}}}}

\newcommand{\Fig}[1]       {Fig.~\ref{#1}}

\newcommand{\Sect}[1]      {Sect.~\ref{#1}}
\newcommand{\Section}[1]   {Section~\ref{#1}}

\newcommand{\Tab}[1]       {Tab.~\ref{#1}}
\newcommand{\Table}[1]     {Table~\ref{#1}}
\newcommand{\todo}[1]      {#1}

\newcommand{\com}[1]       {}
\newcommand{\defrange}     {0.7}
\newcommand{\pur}          {\mathcal{P}}
\iflatexdiff

\renewcommand{\xout}[1]    {\textcolor{red}{\sout{#1}}}
 
\newcommand{\old}[1]       {{\textcolor{red}{\sout{#1}}}}

\else

\renewcommand{\xout}[1]    {}
\newcommand{\old}[1]       {\relax}

\fi

\graphicspath{{./img/}}
\ifdraft
\usepackage{lineno}
\linenumbers
\modulolinenumbers[1]
\usepackage{fancyhdr}
\fi
\begin{document}
\begin{titlepage}
\PHyear{2020}
\PHnumber{210} 
\PHdate{06 November}   
\title{\dtitle}
\ShortTitle{\stitle}
\Collaboration{ALICE Collaboration%
         \thanks{See Appendix~\ref{app:collab} for the list of collaboration members}}
\ShortAuthor{ALICE Collaboration} 
\begin{abstract}
Deuteron production in high-energy collisions is sensitive to the space--time evolution of the collision system, and is typically described by a coalescence mechanism. 
For the first time, we present results on jet-associated deuteron production in \pp\ collisions at $\sqrt{s}=13$ TeV, providing an opportunity to test the established picture for deuteron production in events with a hard scattering.
Using a trigger particle with high transverse-momentum~($\pt>5$~GeV/$c$) as a proxy for the presence of a jet at midrapidity, we observe a measurable population of deuterons being produced around the jet proxy.
The associated deuteron yield measured in a narrow angular range around the trigger particle differs by 2.4--4.8 standard deviations from the uncorrelated background.
The data are described by PYTHIA model calculations featuring baryon coalescence.
\end{abstract}
\end{titlepage}
\newpage
\setcounter{page}{2}
\section{Introduction}
\label{sec:intro}
Measurements of deuterons in high-energy collisions provide insight into baryon production and baryon transport mechanisms which are sensitive to the space--time evolution of the collision system.
Deuteron and anti-deuteron spectra were measured in \pp\ collisions at the CERN ISR~\cite{Alper:1973my,Henning:1977mt} and Tevatron~\cite{Alexopoulos:2000jk}, photo-production processes and deep inelastic scattering of electrons at HERA~\cite{Aktas:2004pq,Chekanov:2007mv}, electron-positron collisions at CLEO~\cite{Asner:2006pw} and LEP~\cite{Schael:2006fd}, and most recently at the LHC in \pp\ collisions at $\sqrt{s}=0.9$, 2.76, 7 and 13~TeV~\cite{Adam:2015vda,Acharya:2017fvb,Acharya:2019rgc,Acharya:2020sfy}, as well as in nucleus--nucleus collisions at SPS~\cite{Anticic:2004yj}, RHIC~\cite{Abelev:2009ae} and LHC~\cite{Adam:2015vda,Acharya:2017dmc,Acharya:2019rys} energies.
Deuteron production can be described by phenomenological models, according to which an (anti-)neutron and (anti-)proton close in phase-space coalesce and bind together~\cite{Hagedorn:1960zz,Sato:1981ez,Gutbrod:1988gt}.
The coalescence mechanism is of broader interest, as it has been employed in describing the production of nuclei and anti-nuclei as large as $^{4}\rm{He}$ and $^{4}\rm{\overline{He}}$~\cite{Agakishiev:2011ib,Acharya:2017bso}, nucleons and hyperons forming hypernuclei~\cite{Adam:2015yta,Adam:2015nca}, searches for exotic states such as pentaquarks~\cite{Karliner:2004hk}, and searches for colorless SUSY-hybrid states with gluinos\com{ such as \~{g}q$\overline{\mbox{q}}'$$^{\pm}$}~\cite{Akers:1995az}.
Statistical hadronization models, which assume particle production in thermal equilibrium, were also successful in explaining the yields of light (anti-)nuclei along with other hadrons in \PbPb\ collisions, but have difficulties to describe the data in smaller systems~\cite{Sharma:2018jqf,Vovchenko:2018fiy}.

New insights may be obtained by studying the production of deuterons from hard processes, which can be explored by their formation within jets. 
To investigate the effects of jets on deuteron production, we employ the two-particle correlation method, as suggested in Ref.~\cite{Sharma:2018dyb}.
Charged particles with transverse momentum ($\pt$) above 5~GeV/$c$ are taken as trigger particles to approximate the jet direction. 
The azimuthal correlation of deuteron candidates with respect to the trigger particle is measured in five $\pt$ intervals between 1 and 4 GeV/$c$.
Impurities are accounted for by using a sideband subtraction method, and deuterons oriented randomly with respect to the trigger particle are subtracted using the \textit{zero yield at minimum}~(ZYAM) method~\cite{Adler:2005ee}.
The integrated yields of associated deuterons obtained within an azimuthal range of $0.7$~rad relative to the trigger particle, representing the region of jet fragmentation, are reported as a function of deuteron $\pt$.
In the coalescence picture, the smaller phase space provided by the jet fragmentation may promote deuteron production.
Hence, the data are compared to model calculations based on PYTHIA~(v8) with a coalescence afterburner~\cite{ALICE-PUBLIC-2017-010}.

The remainder of the letter is organized as follows.
\Section{sec:aliceDet} briefly describes the various ALICE subsystems, the dataset and event selection criteria for the measurement presented.
\Section{sec:ana} discusses the particle identification and correlation analysis methods.
\Section{sec:results} presents the measurement of the associated deuteron yields, discusses the systematic uncertainties, and provides the comparison with the PYTHIA-based coalescence afterburner model.
\Section{sec:conclusions} concludes the letter.
\section{Experimental setup and dataset}
\label{sec:aliceDet}
\label{sec:datasamples}
ALICE is a general purpose detector at the LHC with cylindrical geometry and outer dimensions of $16\times16\times26$~m$^3$~\cite{Aamodt:2008zz}.
A large solenoid magnet provides an uniform magnetic field of $0.5$~T along the beam direction~($z$ direction) and encases the central barrel around the nominal interaction point~(IP) at $z=0$. 
The measurements presented use a subset of the ALICE detector systems, including  the V0~\cite{Abbas:2013taa}, the Inner Tracking System~(ITS)~\cite{Aamodt:2010aa}, the Time Projection Chamber~(TPC)~\cite{Alme:2010ke}, the T0, and the Time-of-Flight (TOF)~\cite{Adam:2016ilk} detectors.
The V0 is a forward detector system used for event triggering.
It consists of two circular planes of plastic scintillators at 87 and 329~cm on opposite sides of the IP covering a pseudorapidity of $-3.7<\eta<-1.7$ and $2.8<\eta<5.1$, respectively.
The ITS is composed of six layers of silicon detectors ranging from 3.9 to 43 cm radius around the beam pipe.
Together with the TPC, it is used for precise reconstruction of the primary vertex position and tracking of charged particles with $\eta<0.9$.
The TPC is a large tracking drift detector~(inner radius $85$~cm, outer radius $250$~cm and length $500$~cm) providing up to 159 space points per track for momentum reconstruction as well as energy loss~(d$E/$d$x$) measurement for particle identification.
The T0 consists of two sets of 12 Cherenkov counters around the beam pipe at $-70$~cm and $374$~cm which provides a measurement of the collision time.
The TOF detector is a cylindrical wall with inner radius 3.7~m from the beam-pipe.
The arrival time of incident hadrons is measured using multi-gap resistive plate chambers with an intrinsic resolution of about $80$~ps.
The particle identification method using a combination of tracking, timing, and energy loss measurements is described in \Sect{sec:ana}.
Further details of the performance of the ALICE detector systems are given in Ref.~\cite{Abelev:2014ffa}.

The analysis is based on the data recorded in \pp\ collisions at $\sqrt{s}=13$ TeV during the years 2015--2018.
The minimum-bias event selection required a hit in both sides of the V0 detector, resulting in approximately 1.8 billion events corresponding to an integrated luminosity of about $30$~nb$^{-1}$.

Additional event selection criteria required at least one track in the ITS with a projection to a vertex position within $0.5$~cm along the beam direction from the position estimated by the T0 collision time.
This requirement suppressed events from out-of-bunch beam background.
The $z$-vertex position was required to be within $10$~cm of the nominal IP to ensure approximately constant $\eta$ acceptance within the detector for all events.
Pile-up events were suppressed by rejecting events with multiple vertices reconstructed by the ITS that are separated by more than 0.8~cm (in the $z$-direction).
Approximately 88\% of the minimum-bias events were accepted for further analysis.

\section{Analysis method}
\label{sec:ana}
Deuteron candidates in several $\pt$ intervals were correlated with charged trigger particles above $5$~GeV/$c$.
The correlation was studied as a function of the azimuthal angle difference~($\dphi$) between deuteron and trigger particle.
In events with multiple triggers and/or deuteron candidates, all combinations were taken into account.
Events with more than one $5$~GeV/$c$ particle correspond to 9.7\% of the selected event sample, while events with more than one deuteron candidate are 0.05\% of the total number of events with a deuteron candidate. 

Deuteron candidates were selected from reconstructed tracks in the central barrel with a pseudorapidity range of $|\eta|<0.9$ that passed several quality criteria.
Tracks were required to contain at least two ITS and 70 TPC clusters, as well as at least 80\% of the maximum possible TPC clusters along its path.
For particle identification, agreement with the expected TOF~(TPC) signal for deuterons within two~(three) standard deviations of the $\pt$-dependent resolution was required, as explained further below.
To suppress secondaries, the distance-of-closest-approach~(DCA) projections of the track to the reconstructed vertex projected on the transverse plane and longitudinal direction, had to be less than $0.5$ and $1$~cm, respectively.

In order to maintain a uniform azimuthal~($\varphi$) distribution for trigger particles, the track quality criteria were relaxed.
In particular, the requirements of having a TOF hit, two ITS clusters, and maximal DCA were not imposed.
The trigger condition $\pt>5$~GeV/$c$ results in an average trigger particle transverse momentum of 6.7~GeV/$c$.

\begin{figure}[t]
   \centering
   \begin{subfigure}[b]{0.32\textwidth}
      \includegraphics[width=\textwidth]{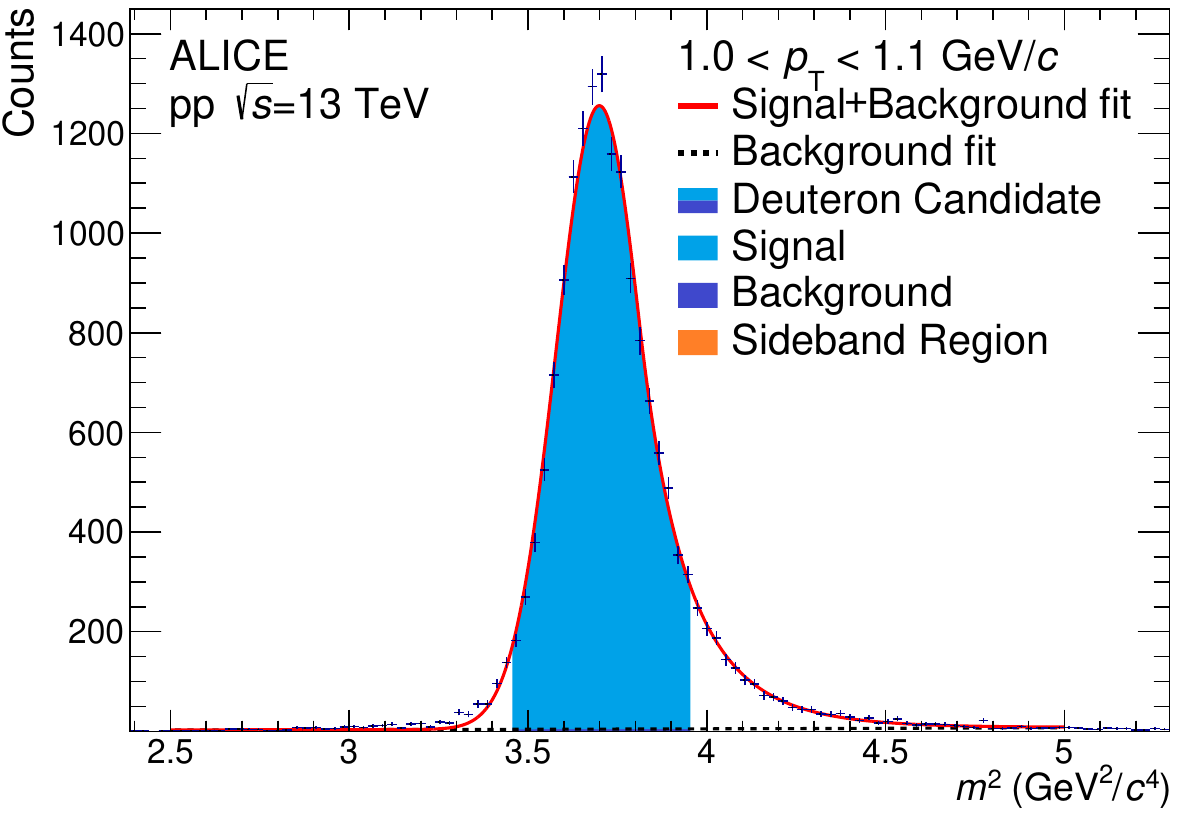}
      \caption{1.0 < \pt~< 1.1 GeV/$c$ interval}
   \end{subfigure} \hspace{1mm}
   \begin{subfigure}[b]{0.32\textwidth}
      \includegraphics[width=\textwidth]{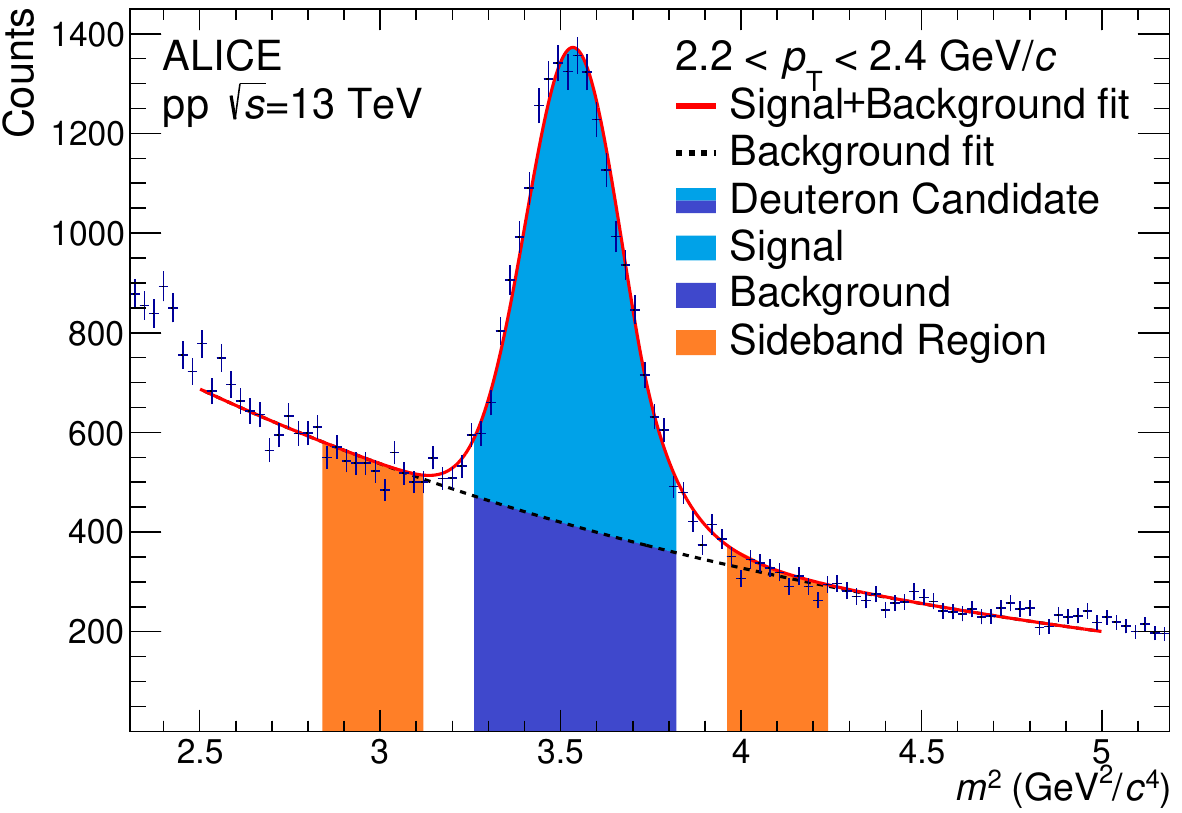}
      \caption{2.2 < \pt~< 2.4 GeV/$c$ interval}
   \end{subfigure} \hspace{1mm}
   \begin{subfigure}[b]{0.32\textwidth}
      \includegraphics[width=\textwidth]{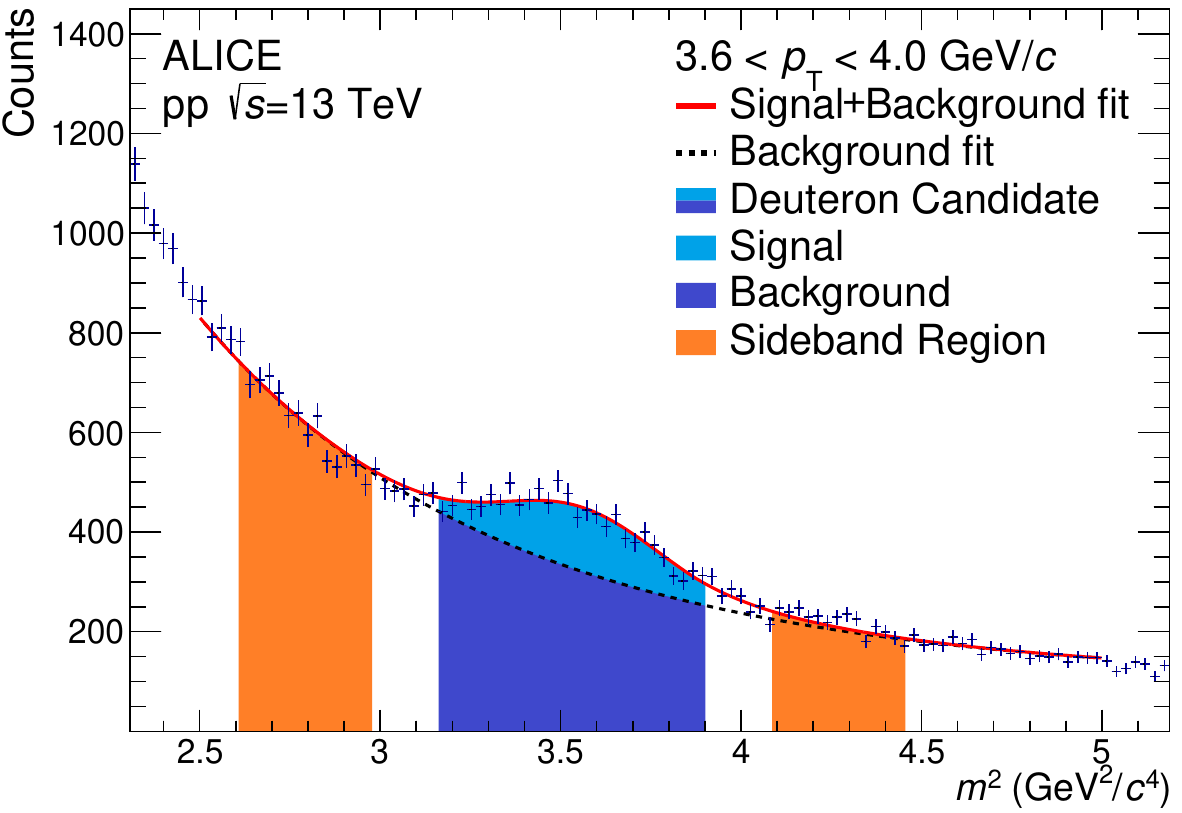}
      \caption{3.6 < \pt~< 4.0 GeV/$c$ interval}
   \end{subfigure}    
   \caption{Example $m^2$-distributions for a)~low,  b)~intermediate and c)~high \pt\ intervals.
            The signal plus background fit is shown as a solid~(red) line, and the extracted background as a dotted~(black) line.
            The $\pm2$~standard deviation candidate region around the mean from the fit is shown in blue. 
            In~(a) no sideband region is visible as the purity is essentially unity.  
            In~(b) and c) the sideband regions are the shaded~(orange) areas between \todo{3--5} standard deviations on both sides of the peak.  
            In the candidate region, the signal is depicted in light blue, while the background is shown in dark blue.
            The purity in the candidate region is approximately 100\% in~(a), 60\% in~(b) and 25\% in~(c).}
   \label{fig:purity}
\end{figure}

The time-of-flight~($t$) of a charged particle was obtained using the difference between the event collision time and the arrival time at the TOF.
Together with the momentum~($p$) and path length~($L$) from the track reconstruction, the mass-squared~($m^2$),
\begin{equation}
 m^2 = \frac{p^2}{c^2}\left(\frac{t^2c^2}{L^2}-1\right)
 \label{equ:mass_squared}
\end{equation}
was calculated for deuteron candidates.
Example $m^2$ distributions of deuteron candidates for different $\pt$ intervals are shown in \Fig{fig:purity}.
The signal component was fit using a Crystal Ball function~\cite{ALICE-PUBLIC-2015-006}.
The standard deviation was approximated by the width of its Gaussian core.
An exponential was used for the background.
An agreement within two standard deviations of the expected $m^2$ value given by the fit\com{ as a function of $\pt$} was required.
Removing candidates with d$E/$d$x$ measured using the TPC outside of three standard deviations from the expected value of deuterons significantly reduced the background, especially in the \pt~region below $2$~GeV/$c$.

The deuteron purity was estimated from integration over the signal and background components of the $m^2$ fit functions.
The purity was measured in fine \pt~intervals and then averaged with statistical weights for the correlation measurement into five intervals, given in \Tab{tab:purity}.
The lower limit of the kinematic range was set to 1 GeV/$c$ to reduce the contamination by secondary~(knock-out) deuterons from spallation in detector material to the percent level~\cite{Adam:2015vda,Acharya:2017fvb}.
The purity is close to 100\% for $\pt\lesssim1.8$ GeV/$c$. 
At larger $\pt$, the background increases gradually and the purity drops to about 25\% in the highest $\pt$ interval.

\begin{table}[t!]
  \centering
   \caption{\label{tab:purity}Deuteron purity estimates for coarse \pt~intervals.}
  \begin{tabular}{l|ccccc}\hline\hline
   \pt-range~(GeV/$c$) &  1.0--1.35    & 1.35--1.8    & 1.8--2.4     & 2.4--3.0     & 3.0--4.0    \\\hline
   Purity~(\%)         &  99.5$\pm$0.1 & 98.4$\pm$0.4 & 75.5$\pm$1.7 & 46.1$\pm$1.9 & 25.5$\pm$1.4 \\\hline\hline
  \end{tabular}\vspace{2mm}
\end{table}

A mixed-event technique was applied to correct for pair efficiency effects caused by non-uniformities of the $\varphi$ acceptance. 
To this end, every deuteron candidate was correlated with 15 trigger particles selected from different events, which were categorized into ten event classes employing five multiplicity and two $z$-vertex intervals. 
The integral of the resulting mixed-event $\dphi$ distribution was normalized to one. 
The raw $\dphi$ distribution of deuteron candidates relative to the trigger particle is divided by the normalized mixed-event distribution, resulting in the ratio $C_{\rm deut.cand.}$. 
The rather small number of events having both the trigger particle and a deuteron candidate did not permit a further separation into intervals of rapidity.
As a result, triggers and deuterons on the edge of the pseudorapidity range~(|$\eta$| < 0.9) have roughly half the probability of being paired compared to those in the central region, an effect that would be corrected for with mixing in two dimensions~\cite{Oh:2016dxe}. 
Depending on the purity~($\pur$) for a given \pt~interval, a fraction of the \dphi\ yield arises from misidentified tracks amongst the deuteron candidates.
The contribution to the yield from misidentified tracks was subtracted using \dphi-correlations obtained in the sideband regions of the $m^2$ distributions with weights from purity estimates, according to 
\begin{equation}
 C_{\rm deuteron}(\dphi) = C_{\rm deut.cand.}(\dphi) - \left(1-\pur\right) \, \frac{N_{\rm deut.cand.}}{N_{\rm sideband}} \, C_{\rm sideband}(\dphi)\,,
 \label{equ:yield}
\end{equation}
\vspace{1mm}
where ${N_{\rm deut.cand.}}/{N_{\rm sideband}}$ was used to normalize the number of associated counts in the sideband region~($C_{\rm sideband}$) to that of the deuteron candidate region~($C_{\rm deut.cand.}$).
The distribution $C_{\rm deuteron}$ represents the correlated yield with respect to \dphi\ between the trigger particle and associated deuterons.
The sideband selection was chosen to be between 3--4 standard deviations on both sides of the peak.
A Monte Carlo simulation,  where (anti-)deuterons were injected into \pp\ events generated by PYTHIA~\cite{Sjostrand:2006za} was used to determine the momentum-dependent tracking efficiency~($\epsilon$) and acceptance~($A$).
Their product strongly rises from 0.2 at $\pt=1$~GeV/$c$ and levels out at about $0.55$ above $1.5$~GeV/$c$.
The corrected deuteron yield per trigger particle~($\Yd$) was then obtained from 
\begin{equation}
 \Yd(\dphi) = \frac{C_{\rm deuteron}(\dphi)}{N_{\rm trig}} \frac{1}{\epsilon\cdot A}\,,
\label{equ:yield3}
\end{equation}
in the five intervals of deuteron $\pt$, where $N_{\rm trig}$ is the total number of trigger particles.
A correction for efficiency and acceptance of the trigger particles, which are approximately constant above $5$~GeV/$c$, was not applied because the related corrections would cancel in the ratio.
The corrected per-trigger yield distributions were obtained independently for deuterons and anti-deuterons\com{ and found to agree within the statistical uncertainties} and then added for the final results.

\section{Results}
\label{sec:results}
\label{sec:sys}
\begin{figure}[t!]
\begin{multicols}{2}
\centering
\includegraphics[width=1.0\linewidth]{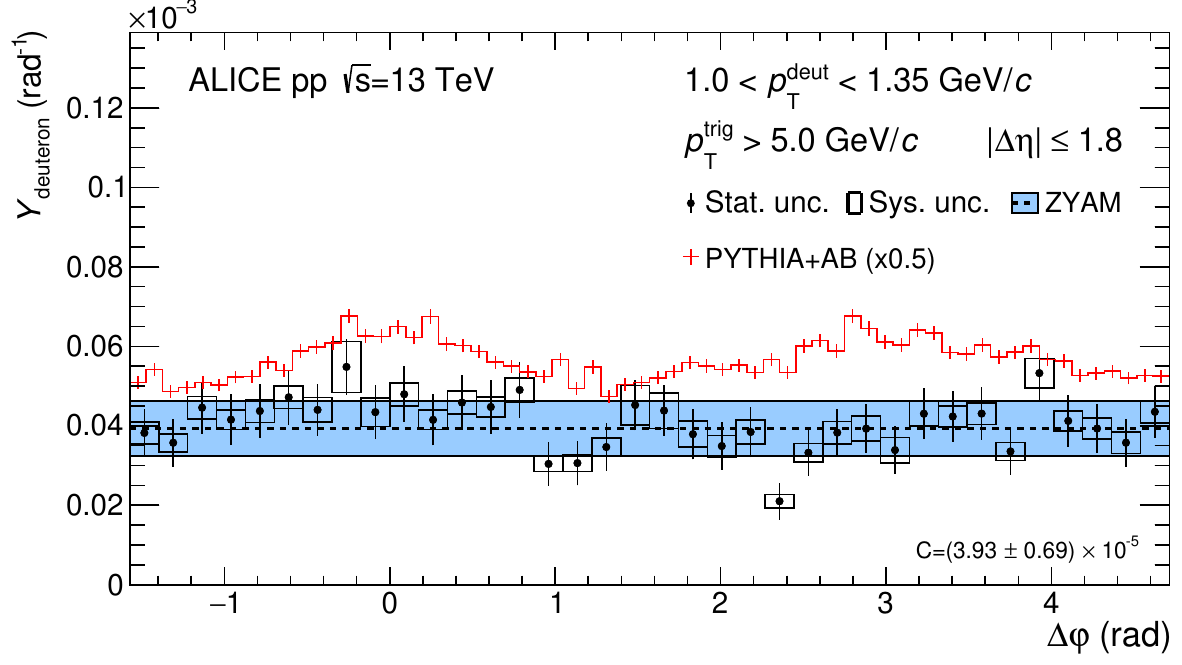}
\vspace{2mm}
\includegraphics[width=1.0\linewidth]{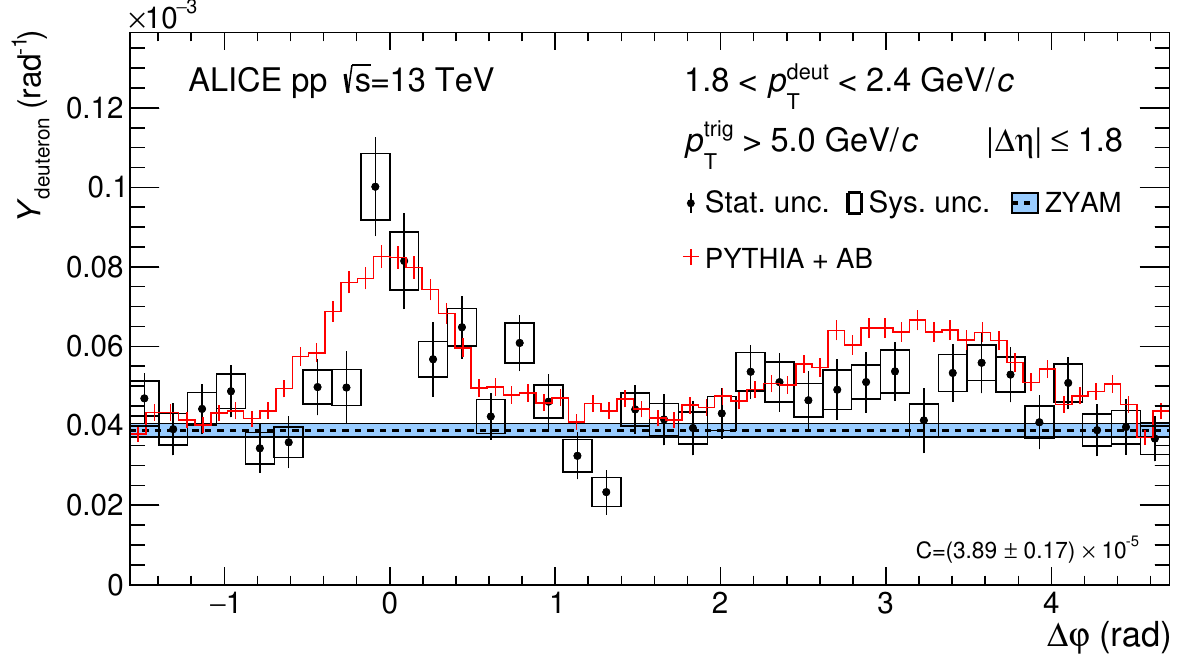}
\vspace{2mm}
\includegraphics[width=1.0\linewidth]{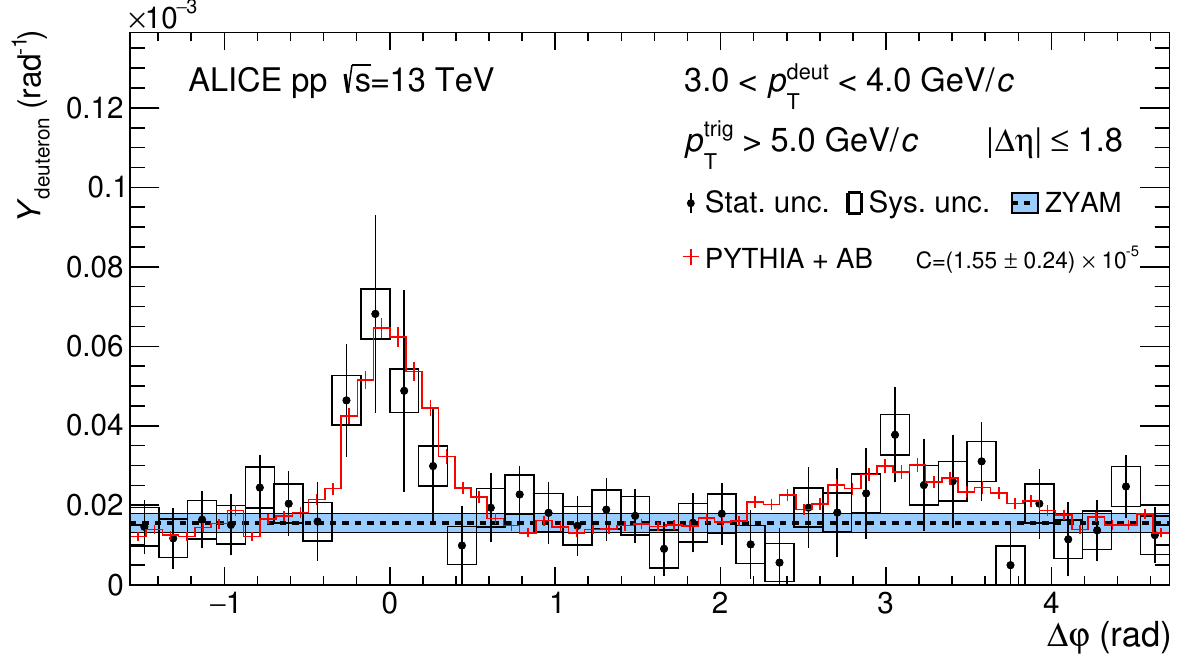}\\
\vspace{2mm}
\includegraphics[width=1.0\linewidth]{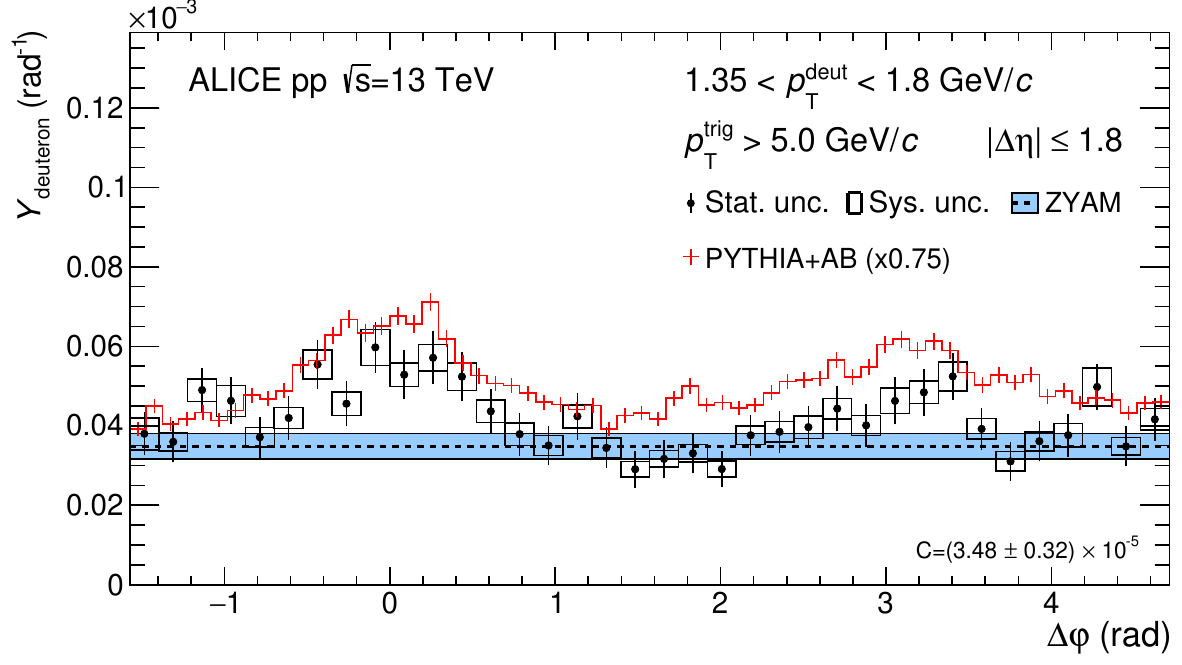}
\vspace{2mm}
\includegraphics[width=1.0\linewidth]{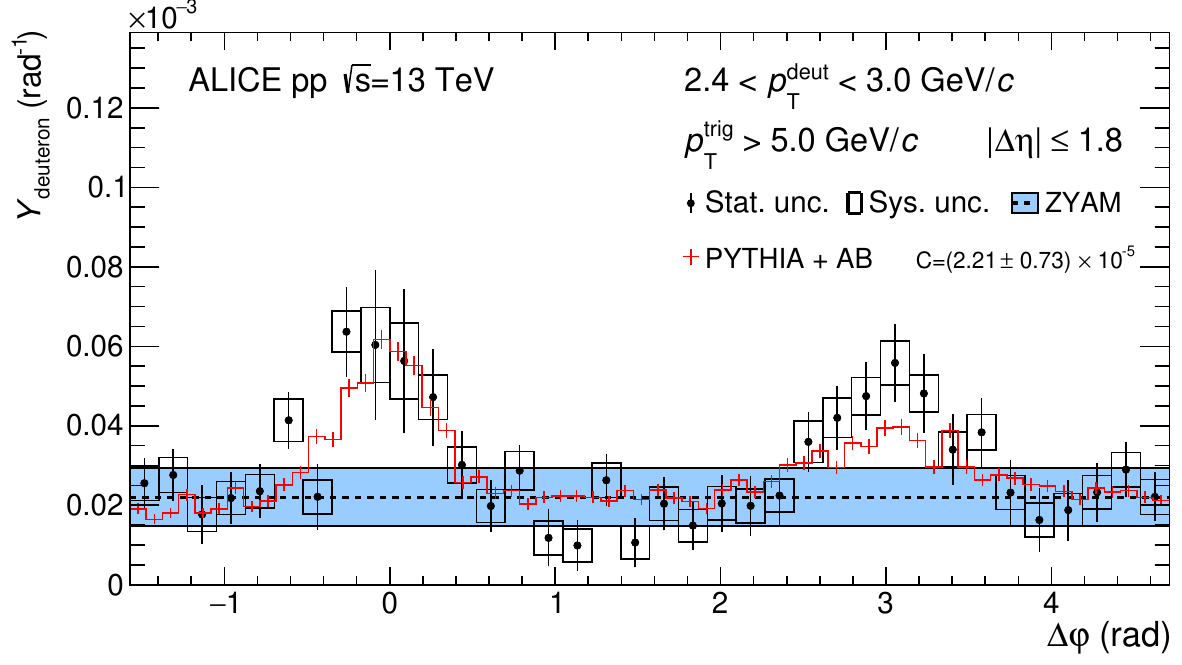}\\
\end{multicols}
\caption{The per-trigger associated yield versus $\Delta\varphi$ for charged particles with $\pt>5.0$ GeV/$c$ and associate deuterons and anti-deuterons for different associate \pt~intervals: 1.0--1.35, 1.35--1.8, 1.8--2.4, 2.4--3.0, and 3.0--4.0 GeV/$c$\com{~(ordered left to right, then down)}.  
         The markers represent the data points with statistical uncertainties, while the boxes represent the systematic uncertainties associated with tracking, purity, and sideband selection. 
         The dotted line shows the ZYAM background estimate and the blue band is the uncertainty associated with the ZYAM estimate.  
         Histogram lines are PYTHIA 8.2 (Monash) model calculations with a coalescence afterburner with $p_0=110$~MeV/$c$.
         The calculation was scaled by $0.5$ and $0.75$ in the first two intervals, required to approximately describe the measured deuteron spectrum at 13 TeV, as explained in the text.}
  \label{fig:correlation}
\end{figure}

The per-trigger associated yield $\Yd$ versus $\dphi$, which represents the probability of deuterons and anti-deuterons being found within a specified \pt~interval and within \dphi\ of a high-\pt~($>5$~GeV/$c$) trigger hadron, is shown in \Fig{fig:correlation} for five deuteron \pt\ intervals.
The markers represent the data points with statistical uncertainties, while the boxes show the total systematic uncertainty. 

Several independent sources of uncertainty associated with tracking, particle identification, sideband correction, and purity, as well as efficiency and acceptance were included into the total systematic uncertainty.
Individual sources were estimated as follows:
a)~the DCA cut was narrowed from 0.5~(1.0)~cm in the $xy$-plane ($z$-axis) to 0.1~(0.1)~cm,
b)~the minimum number of TPC clusters for a track was increased from 70 to 90 hits,
c)~the TOF particle identification requirement on the mass-squared to be within 2 standard deviations of the mean mass was relaxed to 3 standard deviations, 
d)~the mass-squared range used to select the sidebands was changed from 3--4 standard deviations from the mean to 4--5 standard deviations, 
e)~the TPC particle identification requirement of agreement within three standard deviations was tightened to two standard deviations,
f)~the purity calculation from signal and background fit functions was compared to the purity found using bin-counting for the signal and a fit for the background, and
g)~the mixed-event correction in $\dphi$ was not applied.
In addition, a $\dphi$-independent uncertainty of 5\% was applied to account for deficiencies in the deuteron efficiency and acceptance corrections.
A separate purity and track selection efficiency was estimated for each change associated with the deuteron candidate track selection.
The resulting variation~(i.e.\ $p/\epsilon\times A$) was found to differ by less than 10\% from the baseline value obtained using the standard selection.

\Table{tab:sys_uncertainty} summarizes the various systematic uncertainties for the five \pt~intervals.

\begin{table}[t!]
  \centering
    \caption{\label{tab:sys_uncertainty}Uncertainties for each associated \pt~interval.
  \textit{Top:} Statistical uncertainty averaged over all \dphi-intervals.
  \textit{Middle:} Contributions to systematic uncertainties for the different sources described in the text as well as the total, which is obtained from adding the individual contributions in quadrature.
  \textit{Bottom:} Uncertainty associated with the determination of the ZYAM value.
  } 
  \begin{tabular}{l|ccccc}\hline\hline
    \pt-range (GeV/$c$)       &  1.0--1.35   & 1.35--1.8    & 1.8--2.4     & 2.4--3.0     & 3.0--4.0    \\\hline
    Statistical unc.           &  15.6\%      &  13.4\%      &  15.4\%      &  31.7\%      &  57.6\%     \\ \hline
     Sources of sys.\ unc. \\\hline
     a) DCA cut                     &  3.6\%      &  3.5\%      &  2.4\%      &  0.4\%       &  7.6\%     \\
     b) TPC clu.\ min.              &  13.2\%     &  9.7\%      &  0.5\%      &  0.0\%       &  25.2\%    \\
     c) TOF-PID                     &  9.2\%      &  7.3\%      &  17.2\%     &  5.6\%       &  31.8\%    \\
     d) Sidebands                   &  1.9\%      &  0.5\%      &  14.5\%     &  24.8\%      &  14.3\%    \\
     e) TPC-PID                     &  7.0\%      &  2.5\%      &  3.4\%      &  11.2\%      &  20.6\%    \\
     f) Purity det.\                &  0.0\%      &  0.2\%      &  5.0\%      &  11.1\%      &  3.8\%     \\
     g) Mixing                      &  7.7\%      &  11.2\%     &  9.3\%      &  12.7\%      &  5.3\%     \\
     Tracking eff.                  &  5\%        &  5\%        &  5\%        &   5\%        &  5\%       \\\hline
     Total sys.\ unc.               &  20.3\%     &  17.8\%     &  25.7\%     &  32.9\%      &  49.0\%    \\\hline
     ZYAM unc.                      &  101.0\%    &  19.6\%     &  3.7\%      &  27.4\%      &  10.5\%     \\\hline\hline
  \end{tabular}
\end{table}

The resulting systematic uncertainties are largely point-to-point correlated in $\dphi$.
Hence, the shape of the distributions shown in \Fig{fig:correlation} exhibits for all \pt-intervals, except the lowest, a characteristic double-peak structure reminiscent of hard scattering, albeit sitting on a large pedestal value indicative of a large contribution of deuterons produced in the underlying event.
To quantify the per-trigger associated yield of deuterons, the contribution of the uncorrelated background was estimated using the ZYAM method~\cite{Adler:2005ee}.
The ZYAM value was obtained by taking the average over the ranges $\frac{\pi}{2} \pm \frac{\pi}{9}$ and $\frac{3\pi}{2} \pm \frac{\pi}{9}$, which includes eight \dphi\ intervals. 
To estimate the corresponding uncertainty, also reported in \Tab{tab:sys_uncertainty}, we fit a parabola to the $\frac{\pi}{2} \pm \frac{\pi}{9}$ region and use its vertex value as an alternative ZYAM estimate. 
The ZYAM uncertainty, constructed by these two ways, is as such subject to statistical fluctuations. 
The central ZYAM value along with its uncertainty are shown as a band in \Fig{fig:correlation}. 
In the lowest \pt-interval the point-to-point statistical fluctuations in the data are greater in magnitude than the potential underlying trend, resulting in a large ZYAM uncertainty, which demonstrates that the separation between correlated yield and the uncorrelated background is not possible.
In all other \pt-intervals a pronounced jet--associated deuteron enhancement relative to the ZYAM value is visible.

In \Fig{fig:correlation}, the data are also compared to model calculations, based on PYTHIA 8.2 (Monash)~\cite{Sjostrand:2014zea,Skands:2014pea}, including a coalescence afterburner (AB) following~Ref.~\cite{ALICE-PUBLIC-2017-010} for deuteron production, which otherwise is absent in PYTHIA.
In the coalescence model, a (anti-)proton is combined with a (anti-)neutron if each of their momenta in their centre-of-mass frame is smaller than $p_0$, the sole free parameter of the model.
Using $p_0=110$~MeV/$c$, the model describes the deuteron spectra in \pp\ collisions at 7 TeV above $1.5$~GeV/$c$ within uncertainties of about 10\%, while it overpredicts the data by up to 50\% between 1--1.5~GeV/$c$~\cite{Acharya:2017fvb,ALICE-PUBLIC-2017-010}.
Using the same value of $p_0=110$~MeV/$c$, a similar agreement is achieved for the data at 13 TeV~\cite{Acharya:2020sfy}.
The deviations at low $\pt$ of up to 50\% originate from small differences of the level of 10--20\% between the measured and calculation proton yields~\cite{Acharya:2020zji}.
Since there is a large contribution from the underlying event, the calculation in \Fig{fig:correlation} was scaled by 0.5 and 0.75 in the lowest two intervals, to take into account the difference between the model and the data on inclusive deuteron production.
The coalescence model calculation describes the data with the exception of the lowest two associated $\pt$ intervals, where it tends to overpredict the data.

\begin{figure}[t!]
    \centering
    \includegraphics[width=4.5in]{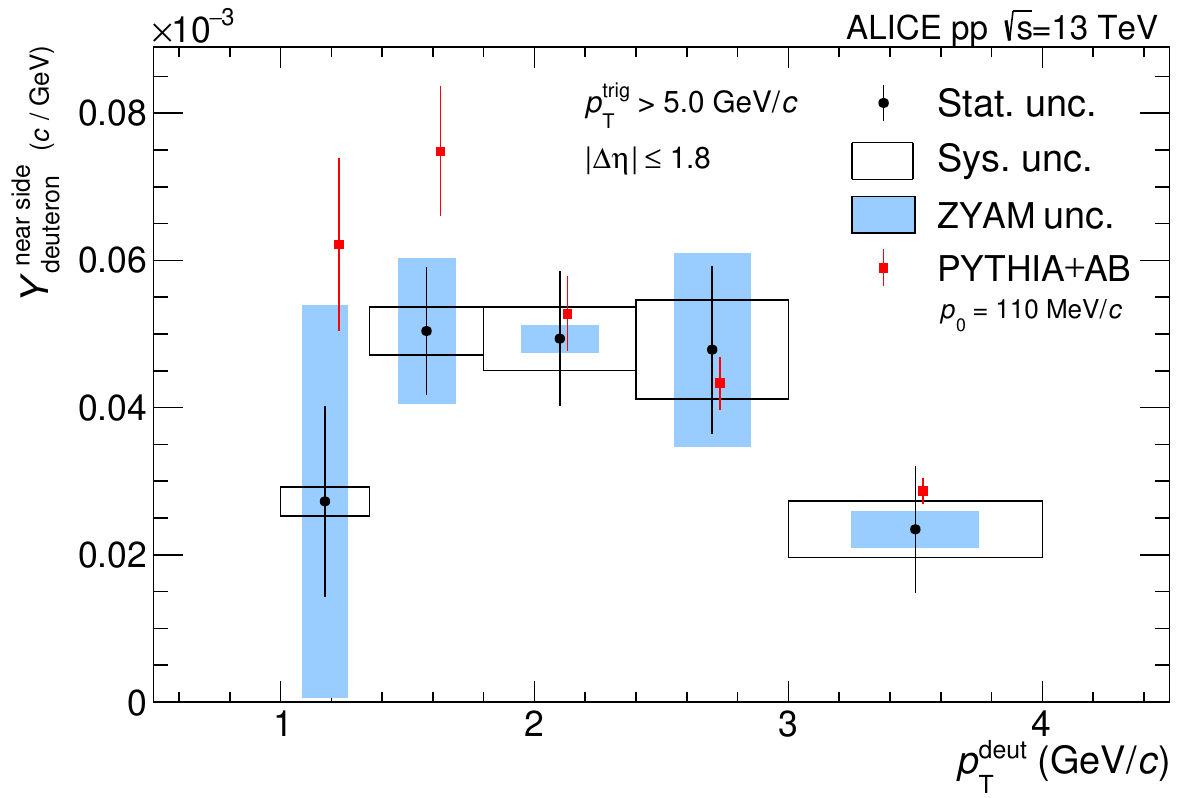}
    \caption{The per-trigger associated-deuteron integrated yield for trigger particles above $5$~GeV/$c$ on the near side versus \pt\ of the associated deuterons and anti-deuterons.
             Vertical bars show statistical uncertainties, open boxes systematic uncertainties, and shaded~(blue) boxes show the uncertainty related to the subtraction of the uncorrelated background using the ZYAM method.
             Square markers are calculations using PYTHIA 8.2 (Monash) with a coalescence afterburner, displaced by 30~MeV/$c$ for better visibility.}
    \label{fig:yields} 
\end{figure}

To extract the per-trigger correlated yield in the jet peak region, $\Yd$ above the ZYAM line is integrated within $|\dphi|<\defrange$~rad,
\begin{equation}
\Yd^{\rm near\ side}=\int_{-\defrange}^{+\defrange} \left(\Yd(\varphi)-C_{\rm ZYAM}\right) \, {\rm d}\varphi\,.
\end{equation}
The per-trigger associated-deuteron integrated yield on the near side as a function of deuteron \pt\ is presented in \Fig{fig:yields}.
The systematic uncertainties from the correlation measurement, which are largely correlated, and from the ZYAM determination, which are largely uncorrelated across deuteron $\pt$, are shown separately. 
For every \pt\ interval except the first, the deuteron yield is between $2.4$ and $4.8$ standard deviations larger than zero~(considering the quadratic sum of statistical, systematic and ZYAM uncertainties), indicating a contribution of deuterons produced in the vicinity of the trigger particle.
The yield of deuterons in the jet peak relative to the production in the underlying event was estimated by computing the ratio of the per trigger yield to the ZYAM value multiplied by $2\pi$. 
The resulting fraction of deuterons produced in the jet is about $8$--$15$\%, increasing with increasing $\pt$, indicating that in the $\pt$ ranges explored by the measurement, the majority of the deuterons are produced in the underlying event.
The model calculations, integrated and corrected using ZYAM in the same way as the data, are in agreement with the data.
The fore-mentioned trend of the calculation to overpredict the data in the two lowest $\pt$ intervals is still present, but  not significant given the large uncertainty from the ZYAM method. 

\section{Conclusions}
\label{sec:conclusions}
Using a high-momentum particle~(\mbox{$\pt>5$}~GeV/$c$) as a proxy for the presence of a jet at midrapidity, we measured the per-trigger yield of associated deuterons and anti-deuterons \com{relative to the azimuthal direction of the trigger particle} in five $\pt$ bins, ranging from 1 to 4~GeV/$c$ in \pp\ collisions at $\sqrt{s}=13$~TeV.
The associated yield integrated within a narrow angular range of the trigger particle is between 2.4 and 4.8 standard deviations above the uncorrelated background in every deuteron $\pt$ interval above $1.35$~GeV/$c$.
In the region of trigger and deuteron \pt\ probed by our measurement, the fraction of deuterons correlated with jets are about 10\% of the number in the underlying event. 
The data are described by PYTHIA model calculations when deuteron production via coalescence is included.

\newenvironment{acknowledgement}{\relax}{\relax}
\begin{acknowledgement}
\section*{Acknowledgments}
\ifdraft
\else

The ALICE Collaboration would like to thank all its engineers and technicians for their invaluable contributions to the construction of the experiment and the CERN accelerator teams for the outstanding performance of the LHC complex.
The ALICE Collaboration gratefully acknowledges the resources and support provided by all Grid centres and the Worldwide LHC Computing Grid (WLCG) collaboration.
The ALICE Collaboration acknowledges the following funding agencies for their support in building and running the ALICE detector:
A. I. Alikhanyan National Science Laboratory (Yerevan Physics Institute) Foundation (ANSL), State Committee of Science and World Federation of Scientists (WFS), Armenia;
Austrian Academy of Sciences, Austrian Science Fund (FWF): [M 2467-N36] and Nationalstiftung f\"{u}r Forschung, Technologie und Entwicklung, Austria;
Ministry of Communications and High Technologies, National Nuclear Research Center, Azerbaijan;
Conselho Nacional de Desenvolvimento Cient\'{\i}fico e Tecnol\'{o}gico (CNPq), Financiadora de Estudos e Projetos (Finep), Funda\c{c}\~{a}o de Amparo \`{a} Pesquisa do Estado de S\~{a}o Paulo (FAPESP) and Universidade Federal do Rio Grande do Sul (UFRGS), Brazil;
Ministry of Education of China (MOEC) , Ministry of Science \& Technology of China (MSTC) and National Natural Science Foundation of China (NSFC), China;
Ministry of Science and Education and Croatian Science Foundation, Croatia;
Centro de Aplicaciones Tecnol\'{o}gicas y Desarrollo Nuclear (CEADEN), Cubaenerg\'{\i}a, Cuba;
Ministry of Education, Youth and Sports of the Czech Republic, Czech Republic;
The Danish Council for Independent Research | Natural Sciences, the VILLUM FONDEN and Danish National Research Foundation (DNRF), Denmark;
Helsinki Institute of Physics (HIP), Finland;
Commissariat \`{a} l'Energie Atomique (CEA) and Institut National de Physique Nucl\'{e}aire et de Physique des Particules (IN2P3) and Centre National de la Recherche Scientifique (CNRS), France;
Bundesministerium f\"{u}r Bildung und Forschung (BMBF) and GSI Helmholtzzentrum f\"{u}r Schwerionenforschung GmbH, Germany;
General Secretariat for Research and Technology, Ministry of Education, Research and Religions, Greece;
National Research, Development and Innovation Office, Hungary;
Department of Atomic Energy Government of India (DAE), Department of Science and Technology, Government of India (DST), University Grants Commission, Government of India (UGC) and Council of Scientific and Industrial Research (CSIR), India;
Indonesian Institute of Science, Indonesia;
Istituto Nazionale di Fisica Nucleare (INFN), Italy;
Institute for Innovative Science and Technology , Nagasaki Institute of Applied Science (IIST), Japanese Ministry of Education, Culture, Sports, Science and Technology (MEXT) and Japan Society for the Promotion of Science (JSPS) KAKENHI, Japan;
Consejo Nacional de Ciencia (CONACYT) y Tecnolog\'{i}a, through Fondo de Cooperaci\'{o}n Internacional en Ciencia y Tecnolog\'{i}a (FONCICYT) and Direcci\'{o}n General de Asuntos del Personal Academico (DGAPA), Mexico;
Nederlandse Organisatie voor Wetenschappelijk Onderzoek (NWO), Netherlands;
The Research Council of Norway, Norway;
Commission on Science and Technology for Sustainable Development in the South (COMSATS), Pakistan;
Pontificia Universidad Cat\'{o}lica del Per\'{u}, Peru;
Ministry of Science and Higher Education, National Science Centre and WUT ID-UB, Poland;
Korea Institute of Science and Technology Information and National Research Foundation of Korea (NRF), Republic of Korea;
Ministry of Education and Scientific Research, Institute of Atomic Physics and Ministry of Research and Innovation and Institute of Atomic Physics, Romania;
Joint Institute for Nuclear Research (JINR), Ministry of Education and Science of the Russian Federation, National Research Centre Kurchatov Institute, Russian Science Foundation and Russian Foundation for Basic Research, Russia;
Ministry of Education, Science, Research and Sport of the Slovak Republic, Slovakia;
National Research Foundation of South Africa, South Africa;
Swedish Research Council (VR) and Knut \& Alice Wallenberg Foundation (KAW), Sweden;
European Organization for Nuclear Research, Switzerland;
Suranaree University of Technology (SUT), National Science and Technology Development Agency (NSDTA) and Office of the Higher Education Commission under NRU project of Thailand, Thailand;
Turkish Atomic Energy Agency (TAEK), Turkey;
National Academy of  Sciences of Ukraine, Ukraine;
Science and Technology Facilities Council (STFC), United Kingdom;
National Science Foundation of the United States of America (NSF) and United States Department of Energy, Office of Nuclear Physics (DOE NP), United States of America.
\fi
\end{acknowledgement}

\bibliographystyle{utphys}
\bibliography{biblio}{}
\newpage
\appendix
\section{The ALICE Collaboration}
\label{app:collab}
\ifdraft
\else
\begingroup
\small
\begin{flushleft}


S.~Acharya$^{\rm 142}$, 
D.~Adamov\'{a}$^{\rm 97}$, 
A.~Adler$^{\rm 75}$, 
J.~Adolfsson$^{\rm 82}$, 
G.~Aglieri Rinella$^{\rm 35}$, 
M.~Agnello$^{\rm 31}$, 
N.~Agrawal$^{\rm 55}$, 
Z.~Ahammed$^{\rm 142}$, 
S.~Ahmad$^{\rm 16}$, 
S.U.~Ahn$^{\rm 77}$
% ...

\section*{Affiliation notes}

$^{\rm I}$ Deceased\\
$^{\rm II}$ Also at: Italian National Agency for New Technologies, Energy and Sustainable Economic Development (ENEA), Bologna, Italy\\
$^{\rm III}$ Also at: Dipartimento DET del Politecnico di Torino, Turin, Italy\\
$^{\rm IV}$ Also at: M.V. Lomonosov Moscow State University, D.V. Skobeltsyn Institute of Nuclear, Physics, Moscow, Russia\\
$^{\rm V}$ Also at: Institute of Theoretical Physics, University of Wroclaw, Poland\\

\section*{Collaboration Institutes}

$^{1}$ A.I. Alikhanyan National Science Laboratory (Yerevan Physics Institute) Foundation, Yerevan, Armenia\\
$^{2}$ AGH University of Science and Technology, Cracow, Poland\\
$^{3}$ Bogolyubov Institute for Theoretical Physics, National Academy of Sciences of Ukraine, Kiev, Ukraine\\
$^{4}$ Bose Institute, Department of Physics  and Centre for Astroparticle Physics and Space Science (CAPSS), Kolkata, India\\
$^{5}$ Budker Institute for Nuclear Physics, Novosibirsk, Russia\\
$^{6}$ California Polytechnic State University, San Luis Obispo, California, United States\\
$^{7}$ Central China Normal University, Wuhan, China\\
$^{8}$ Centro de Aplicaciones Tecnol\'{o}gicas y Desarrollo Nuclear (CEADEN), Havana, Cuba\\
$^{9}$ Centro de Investigaci\'{o}n y de Estudios Avanzados (CINVESTAV), Mexico City and M\'{e}rida, Mexico\\
$^{10}$ Chicago State University, Chicago, Illinois, United States\\
$^{11}$ China Institute of Atomic Energy, Beijing, China\\
$^{12}$ Chungbuk National University, Cheongju, Republic of Korea\\
$^{13}$ Comenius University Bratislava, Faculty of Mathematics, Physics and Informatics, Bratislava, Slovakia\\
$^{14}$ COMSATS University Islamabad, Islamabad, Pakistan\\
$^{15}$ Creighton University, Omaha, Nebraska, United States\\
$^{16}$ Department of Physics, Aligarh Muslim University, Aligarh, India\\
$^{17}$ Department of Physics, Pusan National University, Pusan, Republic of Korea\\
$^{18}$ Department of Physics, Sejong University, Seoul, Republic of Korea\\
$^{19}$ Department of Physics, University of California, Berkeley, California, United States\\
$^{20}$ Department of Physics, University of Oslo, Oslo, Norway\\
$^{21}$ Department of Physics and Technology, University of Bergen, Bergen, Norway\\
$^{22}$ Dipartimento di Fisica dell'Universit\`{a} 'La Sapienza' and Sezione INFN, Rome, Italy\\
$^{23}$ Dipartimento di Fisica dell'Universit\`{a} and Sezione INFN, Cagliari, Italy\\
$^{24}$ Dipartimento di Fisica dell'Universit\`{a} and Sezione INFN, Trieste, Italy\\
$^{25}$ Dipartimento di Fisica dell'Universit\`{a} and Sezione INFN, Turin, Italy\\
$^{26}$ Dipartimento di Fisica e Astronomia dell'Universit\`{a} and Sezione INFN, Bologna, Italy\\
$^{27}$ Dipartimento di Fisica e Astronomia dell'Universit\`{a} and Sezione INFN, Catania, Italy\\
$^{28}$ Dipartimento di Fisica e Astronomia dell'Universit\`{a} and Sezione INFN, Padova, Italy\\
$^{29}$ Dipartimento di Fisica e Nucleare e Teorica, Universit\`{a} di Pavia  and Sezione INFN, Pavia, Italy\\
$^{30}$ Dipartimento di Fisica `E.R.~Caianiello' dell'Universit\`{a} and Gruppo Collegato INFN, Salerno, Italy\\
$^{31}$ Dipartimento DISAT del Politecnico and Sezione INFN, Turin, Italy\\
$^{32}$ Dipartimento di Scienze e Innovazione Tecnologica dell'Universit\`{a} del Piemonte Orientale and INFN Sezione di Torino, Alessandria, Italy\\
$^{33}$ Dipartimento di Scienze MIFT, Universit\`{a} di Messina, Messina, Italy\\
$^{34}$ Dipartimento Interateneo di Fisica `M.~Merlin' and Sezione INFN, Bari, Italy\\
$^{35}$ European Organization for Nuclear Research (CERN), Geneva, Switzerland\\
$^{36}$ Faculty of Electrical Engineering, Mechanical Engineering and Naval Architecture, University of Split, Split, Croatia\\
$^{37}$ Faculty of Engineering and Science, Western Norway University of Applied Sciences, Bergen, Norway\\
$^{38}$ Faculty of Nuclear Sciences and Physical Engineering, Czech Technical University in Prague, Prague, Czech Republic\\
$^{39}$ Faculty of Science, P.J.~\v{S}af\'{a}rik University, Ko\v{s}ice, Slovakia\\
$^{40}$ Frankfurt Institute for Advanced Studies, Johann Wolfgang Goethe-Universit\"{a}t Frankfurt, Frankfurt, Germany\\
$^{41}$ Fudan University, Shanghai, China\\
$^{42}$ Gangneung-Wonju National University, Gangneung, Republic of Korea\\
$^{43}$ Gauhati University, Department of Physics, Guwahati, India\\
$^{44}$ Helmholtz-Institut f\"{u}r Strahlen- und Kernphysik, Rheinische Friedrich-Wilhelms-Universit\"{a}t Bonn, Bonn, Germany\\
$^{45}$ Helsinki Institute of Physics (HIP), Helsinki, Finland\\
$^{46}$ High Energy Physics Group,  Universidad Aut\'{o}noma de Puebla, Puebla, Mexico\\
$^{47}$ Hiroshima University, Hiroshima, Japan\\
$^{48}$ Hochschule Worms, Zentrum  f\"{u}r Technologietransfer und Telekommunikation (ZTT), Worms, Germany\\
$^{49}$ Horia Hulubei National Institute of Physics and Nuclear Engineering, Bucharest, Romania\\
$^{50}$ Indian Institute of Technology Bombay (IIT), Mumbai, India\\
$^{51}$ Indian Institute of Technology Indore, Indore, India\\
$^{52}$ Indonesian Institute of Sciences, Jakarta, Indonesia\\
$^{53}$ INFN, Laboratori Nazionali di Frascati, Frascati, Italy\\
$^{54}$ INFN, Sezione di Bari, Bari, Italy\\
$^{55}$ INFN, Sezione di Bologna, Bologna, Italy\\
$^{56}$ INFN, Sezione di Cagliari, Cagliari, Italy\\
$^{57}$ INFN, Sezione di Catania, Catania, Italy\\
$^{58}$ INFN, Sezione di Padova, Padova, Italy\\
$^{59}$ INFN, Sezione di Roma, Rome, Italy\\
$^{60}$ INFN, Sezione di Torino, Turin, Italy\\
$^{61}$ INFN, Sezione di Trieste, Trieste, Italy\\
$^{62}$ Inha University, Incheon, Republic of Korea\\
$^{63}$ Institute for Gravitational and Subatomic Physics (GRASP), Utrecht University/Nikhef, Utrecht, Netherlands\\
$^{64}$ Institute for Nuclear Research, Academy of Sciences, Moscow, Russia\\
$^{65}$ Institute of Experimental Physics, Slovak Academy of Sciences, Ko\v{s}ice, Slovakia\\
$^{66}$ Institute of Physics, Homi Bhabha National Institute, Bhubaneswar, India\\
$^{67}$ Institute of Physics of the Czech Academy of Sciences, Prague, Czech Republic\\
$^{68}$ Institute of Space Science (ISS), Bucharest, Romania\\
$^{69}$ Institut f\"{u}r Kernphysik, Johann Wolfgang Goethe-Universit\"{a}t Frankfurt, Frankfurt, Germany\\
$^{70}$ Instituto de Ciencias Nucleares, Universidad Nacional Aut\'{o}noma de M\'{e}xico, Mexico City, Mexico\\
$^{71}$ Instituto de F\'{i}sica, Universidade Federal do Rio Grande do Sul (UFRGS), Porto Alegre, Brazil\\
$^{72}$ Instituto de F\'{\i}sica, Universidad Nacional Aut\'{o}noma de M\'{e}xico, Mexico City, Mexico\\
$^{73}$ iThemba LABS, National Research Foundation, Somerset West, South Africa\\
$^{74}$ Jeonbuk National University, Jeonju, Republic of Korea\\
$^{75}$ Johann-Wolfgang-Goethe Universit\"{a}t Frankfurt Institut f\"{u}r Informatik, Fachbereich Informatik und Mathematik, Frankfurt, Germany\\
$^{76}$ Joint Institute for Nuclear Research (JINR), Dubna, Russia\\
$^{77}$ Korea Institute of Science and Technology Information, Daejeon, Republic of Korea\\
$^{78}$ KTO Karatay University, Konya, Turkey\\
$^{79}$ Laboratoire de Physique des 2 Infinis, Ir\`{e}ne Joliot-Curie, Orsay, France\\
$^{80}$ Laboratoire de Physique Subatomique et de Cosmologie, Universit\'{e} Grenoble-Alpes, CNRS-IN2P3, Grenoble, France\\
$^{81}$ Lawrence Berkeley National Laboratory, Berkeley, California, United States\\
$^{82}$ Lund University Department of Physics, Division of Particle Physics, Lund, Sweden\\
$^{83}$ Moscow Institute for Physics and Technology, Moscow, Russia\\
$^{84}$ Nagasaki Institute of Applied Science, Nagasaki, Japan\\
$^{85}$ Nara Women{'}s University (NWU), Nara, Japan\\
$^{86}$ National and Kapodistrian University of Athens, School of Science, Department of Physics , Athens, Greece\\
$^{87}$ National Centre for Nuclear Research, Warsaw, Poland\\
$^{88}$ National Institute of Science Education and Research, Homi Bhabha National Institute, Jatni, India\\
$^{89}$ National Nuclear Research Center, Baku, Azerbaijan\\
$^{90}$ National Research Centre Kurchatov Institute, Moscow, Russia\\
$^{91}$ Niels Bohr Institute, University of Copenhagen, Copenhagen, Denmark\\
$^{92}$ Nikhef, National institute for subatomic physics, Amsterdam, Netherlands\\
$^{93}$ NRC Kurchatov Institute IHEP, Protvino, Russia\\
$^{94}$ NRC \guillemotleft Kurchatov\guillemotright  Institute - ITEP, Moscow, Russia\\
$^{95}$ NRNU Moscow Engineering Physics Institute, Moscow, Russia\\
$^{96}$ Nuclear Physics Group, STFC Daresbury Laboratory, Daresbury, United Kingdom\\
$^{97}$ Nuclear Physics Institute of the Czech Academy of Sciences, \v{R}e\v{z} u Prahy, Czech Republic\\
$^{98}$ Oak Ridge National Laboratory, Oak Ridge, Tennessee, United States\\
$^{99}$ Ohio State University, Columbus, Ohio, United States\\
$^{100}$ Petersburg Nuclear Physics Institute, Gatchina, Russia\\
$^{101}$ Physics department, Faculty of science, University of Zagreb, Zagreb, Croatia\\
$^{102}$ Physics Department, Panjab University, Chandigarh, India\\
$^{103}$ Physics Department, University of Jammu, Jammu, India\\
$^{104}$ Physics Department, University of Rajasthan, Jaipur, India\\
$^{105}$ Physikalisches Institut, Eberhard-Karls-Universit\"{a}t T\"{u}bingen, T\"{u}bingen, Germany\\
$^{106}$ Physikalisches Institut, Ruprecht-Karls-Universit\"{a}t Heidelberg, Heidelberg, Germany\\
$^{107}$ Physik Department, Technische Universit\"{a}t M\"{u}nchen, Munich, Germany\\
$^{108}$ Politecnico di Bari and Sezione INFN, Bari, Italy\\
$^{109}$ Research Division and ExtreMe Matter Institute EMMI, GSI Helmholtzzentrum f\"ur Schwerionenforschung GmbH, Darmstadt, Germany\\
$^{110}$ Rudjer Bo\v{s}kovi\'{c} Institute, Zagreb, Croatia\\
$^{111}$ Russian Federal Nuclear Center (VNIIEF), Sarov, Russia\\
$^{112}$ Saha Institute of Nuclear Physics, Homi Bhabha National Institute, Kolkata, India\\
$^{113}$ School of Physics and Astronomy, University of Birmingham, Birmingham, United Kingdom\\
$^{114}$ Secci\'{o}n F\'{\i}sica, Departamento de Ciencias, Pontificia Universidad Cat\'{o}lica del Per\'{u}, Lima, Peru\\
$^{115}$ St. Petersburg State University, St. Petersburg, Russia\\
$^{116}$ Stefan Meyer Institut f\"{u}r Subatomare Physik (SMI), Vienna, Austria\\
$^{117}$ SUBATECH, IMT Atlantique, Universit\'{e} de Nantes, CNRS-IN2P3, Nantes, France\\
$^{118}$ Suranaree University of Technology, Nakhon Ratchasima, Thailand\\
$^{119}$ Technical University of Ko\v{s}ice, Ko\v{s}ice, Slovakia\\
$^{120}$ The Henryk Niewodniczanski Institute of Nuclear Physics, Polish Academy of Sciences, Cracow, Poland\\
$^{121}$ The University of Texas at Austin, Austin, Texas, United States\\
$^{122}$ Universidad Aut\'{o}noma de Sinaloa, Culiac\'{a}n, Mexico\\
$^{123}$ Universidade de S\~{a}o Paulo (USP), S\~{a}o Paulo, Brazil\\
$^{124}$ Universidade Estadual de Campinas (UNICAMP), Campinas, Brazil\\
$^{125}$ Universidade Federal do ABC, Santo Andre, Brazil\\
$^{126}$ University of Cape Town, Cape Town, South Africa\\
$^{127}$ University of Houston, Houston, Texas, United States\\
$^{128}$ University of Jyv\"{a}skyl\"{a}, Jyv\"{a}skyl\"{a}, Finland\\
$^{129}$ University of Liverpool, Liverpool, United Kingdom\\
$^{130}$ University of Science and Technology of China, Hefei, China\\
$^{131}$ University of South-Eastern Norway, Tonsberg, Norway\\
$^{132}$ University of Tennessee, Knoxville, Tennessee, United States\\
$^{133}$ University of the Witwatersrand, Johannesburg, South Africa\\
$^{134}$ University of Tokyo, Tokyo, Japan\\
$^{135}$ University of Tsukuba, Tsukuba, Japan\\
$^{136}$ Universit\'{e} Clermont Auvergne, CNRS/IN2P3, LPC, Clermont-Ferrand, France\\
$^{137}$ Universit\'{e} de Lyon, CNRS/IN2P3, Institut de Physique des 2 Infinis de Lyon , Lyon, France\\
$^{138}$ Universit\'{e} de Strasbourg, CNRS, IPHC UMR 7178, F-67000 Strasbourg, France, Strasbourg, France\\
$^{139}$ Universit\'{e} Paris-Saclay Centre d'Etudes de Saclay (CEA), IRFU, D\'{e}partment de Physique Nucl\'{e}aire (DPhN), Saclay, France\\
$^{140}$ Universit\`{a} degli Studi di Foggia, Foggia, Italy\\
$^{141}$ Universit\`{a} di Brescia and Sezione INFN, Brescia, Italy\\
$^{142}$ Variable Energy Cyclotron Centre, Homi Bhabha National Institute, Kolkata, India\\
$^{143}$ Warsaw University of Technology, Warsaw, Poland\\
$^{144}$ Wayne State University, Detroit, Michigan, United States\\
$^{145}$ Westf\"{a}lische Wilhelms-Universit\"{a}t M\"{u}nster, Institut f\"{u}r Kernphysik, M\"{u}nster, Germany\\
$^{146}$ Wigner Research Centre for Physics, Budapest, Hungary\\
$^{147}$ Yale University, New Haven, Connecticut, United States\\
$^{148}$ Yonsei University, Seoul, Republic of Korea\\

\endgroup

\fi
\end{document}